\UseRawInputEncoding
\documentclass[preprint,superscriptaddress,floatfix]{revtex4-2}

\usepackage{graphicx}% Include figure files
\usepackage{dcolumn}% Align table columns on decimal point
\usepackage{verbatim}
\usepackage[colorlinks=true]{hyperref}% add hypertext capabilities
\usepackage{multirow}
\usepackage{bm}% bold math
\usepackage{times}
\usepackage{color}
\usepackage{amstext}
\usepackage{amsmath}
\usepackage{amssymb} 
\usepackage{amsfonts}
\usepackage[final]{microtype}
\usepackage{csquotes,lmodern}
\usepackage{hyperref}
\hypersetup{hidelinks}
\newcommand{\CRO}{Ca$_2$RuO$_4$}
\newcommand{\CROb}{Ca$_3$Ru$_2$O$_7$}
\newcommand{\CRTO}{Ca$_3$(Ru$_{0.99}$Ti$_{0.01}$)$_2$O$_7$}
\newcommand{\CRTOx}{Ca$_3$(Ru$_{1-x}$Ti$_x$)$_2$O$_7$}

\begin{document}

%\title{Photosensitive electronic phase instability and correlations in Ca\ci{3}(Ru\ci{0.99}Ti\ci{0.01})\ci{2}O\ci{7}}
\title{Photoinduced phase switching at a Mott insulator-to-metal transition}  

%% Notice placement of commas and superscripts and use of &
%% in the author list

\author{K.~S.~Rabinovich}
\affiliation{
Max Planck Institute for Solid State Research, Heisenbergstra{\ss}e~1,
70569 Stuttgart, Germany}

\author{A.~N.~Yaresko}
\affiliation{
Max Planck Institute for Solid State Research, Heisenbergstra{\ss}e~1,
70569 Stuttgart, Germany}

\author{R.~D.~Dawson}
\affiliation{
Max Planck Institute for Solid State Research, Heisenbergstra{\ss}e~1,
70569 Stuttgart, Germany}

\author{M.~J.~Krautloher}
\affiliation{
Max Planck Institute for Solid State Research, Heisenbergstra{\ss}e~1,
70569 Stuttgart, Germany}

\author{T.~Priessnitz}
\affiliation{
Max Planck Institute for Solid State Research, Heisenbergstra{\ss}e~1,
70569 Stuttgart, Germany}

\author{Y.-L.~Mathis}
\affiliation{
Institute for Beam Physics and Technology, Karlsruhe Institute of
Technology, 76344 Eggenstein - Leopoldshafen, Germany}

\author{B.~Keimer}
\affiliation{
Max Planck Institute for Solid State Research, Heisenbergstra{\ss}e~1,
70569 Stuttgart, Germany}

\author{A.~V.~Boris} 
\email{A.Boris@fkf.mpg.de}
\affiliation{
Max Planck Institute for Solid State Research, Heisenbergstra{\ss}e~1,
70569 Stuttgart, Germany}

%\linenumbers

\begin{abstract}
Achieving fundamental understanding of insulator-to-metal transitions (IMTs) in strongly correlated systems \cite{Imada1998} and their persistent and reversible control via nonequilibrium drive \cite{Basov2017,Bao2021} are prime targets of current condensed matter research. Photoinduced switching between competing orders in correlated insulators requires a free-energy landscape with nearly degenerate ground states, which is commonly reached through doping, strain, or static electric field \cite{Fiebig1998,Zhang2016}. The associated spatial inhomogeneity leads to a photoinduced phase transition (PIPT) that remains confined near the illuminated region. Here we report
optical spectroscopy experiments at the first-order IMT in the $4d$-electron
compound \CRTO \ and show that specific Ru $d$-$d$ interband transitions
excited by light with a threshold fluence corresponding to the planar density
of Ru atoms can trigger reversible, avalanche-like coherent propagation of
phase interfaces across the full extent of a macroscopic sample, in the absence
of assisting external stimuli. Based on detailed comparison of spectroscopic
data to density functional calculations, we attribute the extraordinary photo-sensitivity
of the IMT to an exceptionally shallow free-energy landscape generated by
the confluence of electron-electron and electron-lattice interactions. Our
findings suggest \CRTO \ as an ideal model system for building and testing
a theory of Mott transition dynamics in the presence of strong electron-lattice
coupling and may pave the way towards nanoscale devices with quantum-level
photosensitivity.
\end{abstract}

\maketitle
\newpage

Light control over phase transitions in perovskite transition metal oxides requires precise tuning of the intrinsic collective instabilities involving spin, charge, orbital, and lattice degrees of freedom. Rotations and distortions of the octahedral network give rise to a delicate interplay between the underlying electronic, magnetic, and crystal structure. Competition between these different degrees of freedom has been extensively studied by ultrafast pump-probe spectroscopy, which has revealed short-lived, transient electronic and magnetic states that are not thermally accessible in equilibrium \cite{Torre2021,Zhang2014}. Only the colossal magnetoresistive manganites have been observed to exhibit long-lived and reversible light-induced phase switching between antiferromagnetic (AFM) charge-ordered insulating and ferromagnetic (FM) metallic states \cite{Fiebig1998,Zhang2016,McLeod2020,Takubo2005}. However, this transition from localized spin to itinerant electron behavior requires an assisting external stimulus, such as a static electric field or epitaxial strain, and remains spatially confined to the illuminated region. Magnetoelastic effects are essential to stabilize AFM insulator and FM metallic phases in a nearly degenerate configuration.

The high level of stability of the charge-ordered insulating phase in manganites is a consequence of the strong tendency of its $3d$ valence electrons to localize as a result of the high ratio of Coulomb interaction to bandwidth. This propensity for electrons to localize is reduced in $4d$-electron materials, particularly ruthenates. This compound family thus offers an auspicious path toward achieving robust light-sensitive phase control, owing to the delicate balance between the competing energy scales of collective instabilities, electronic correlations, and enhanced spin-orbit coupling. Since $4d$ orbitals are more extended, electron correlations are reduced and the ligand field strength is increased. Due to the large crystal field splitting, the $\rm Ru^{4+}$ ions host a low-spin $S=1$ state with four electrons in the $t_{2g}$ manifold, such that the spin-orbit interaction enters as an important
energy scale together with Coulomb repulsion and Hunds coupling. The resulting flattened energy landscape leads to a diverse array of exotic ground states where small light-induced perturbations are expected to cause switching between different phases.

Such extreme sensitivity drives ruthenates, in particular, to exhibit a high degree of susceptibility to isovalent substitution of Ru ions. Here, we focus on the \CROb \ bilayer perovskite system whose ground state is a polar metal \cite{Cao1997,Lee2007,Markovic2020,Sokolov2019,Lei2018,Horio2021,Bertinshaw2021}, in contrast to its Mott insulating single-layer counterpart \CRO \ \cite{Jain2017,Sutter2017,Gretarsson2019}. Dilute substitution of Ru with 1\% Ti in \CRTOx \ restores the insulating
ground state \cite{Ke2011,Tsuda2013,Krautloher2018}. At room temperature the structural, magnetic, and electronic transport properties
of \CRTO \ do not differ appreciably from those of pristine \CROb. Both are
paramagnetic metals (PM-M) and adopt an orthorhombic crystal structure with polar space group $Bb2_1m$. However, upon cooling the two compounds display strikingly different behavior. \CROb \ experiences a pair of consecutive transitions in which AFM-$a$ magnetic ordering at $T_N$ = 56 K (FM (AFM) within (between) bilayers) is followed by a metamagnetic spin reorientation transition into the AFM-$b$ phase that occurs concomitantly with an isostructural transition ($T_s$ = 48 K) \cite{Bao2008,Sokolov2019,Bertinshaw2021}. Despite the opening of the pseudogap due to Fermi surface reconstruction at $T_s$, the material remains a polar metal where itinerant electrons persist within the ferromagnetically ordered bilayers \cite{Lee2007,Markovic2020,Horio2021}. On the other hand, \CRTO \ exhibits a single isostructural Mott transition at $T_{IMT}$ = 55 K from itinerant to localized electronic behavior with the $\rm Ru$ magnetic moments aligned antiferromagnetically (G-AFM-I) within the bilayers.

\section*{Photoinduced insulator to metal switching}

The central issue of this report is our observation of photoinduced phase
switching in the near-threshold behavior of \CRTO. The temperature driven
phase transition at $T_{IMT}$ is first order and characterized by hysteresis
and phase separation at the transition. We use conventional far-field spectroscopy to monitor the hysteretic behavior near the transition point by combining reflected light microscopy (Fig. 1a) with scanning the dielectric permittivity $\varepsilon_1$ at $\hbar \omega = 0.6$ eV during cooling and heating cycles of the sample (Fig. 1b). We find a direct correspondence between the contrast of stripe domains (Fig. 1c) and the in-plane permittivity values of the G-AFM-I ($\varepsilon_1\approx 10$) and PM-M ($\varepsilon_1\approx -6$) phases coexisting within the hysteresis loop. Whereas only the G-AFM-I
and PM-M phases stabilize at slow cooling ($\lesssim 3 $ K/min) where thermodynamic
equilibrium is maintained, additional dark contrast is detected when the
temperature is cycled at a fast cooling rate ($\simeq 5.2$ K/min). We suggest
that the regions of strong dark contrast represent possible transient trapping
of a metastable FM bilayer metallic phase (AFM-$a$ or $b$), consistent with the near-degeneracy of the two types of magnetic order discussed below. The static stripe pattern is stabilized by temperature at
any point within the hysteresis loop, and by applying low fluence light the
sample displays complete switching into the dark contrast PM-M state.

\begin{figure*}[t]
\includegraphics[width=16cm]{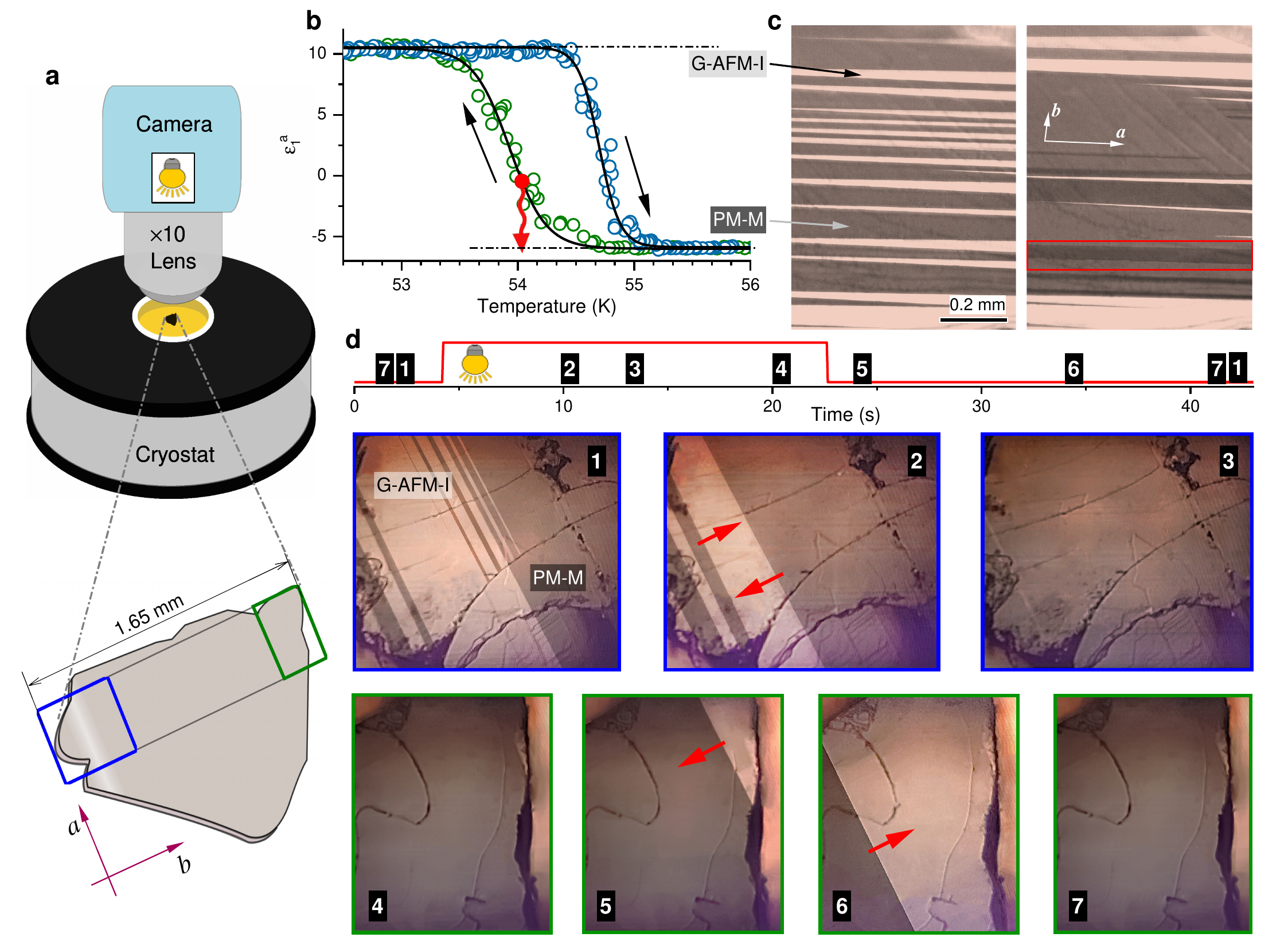}
\caption{\scriptsize{\bf Stripe phase and photoinduced Mott insulator-to-metal
transition in \CRTO.
\\ }
\scriptsize
{\bf a}, Schematic of the reflected light microscopy setup, equipped with
a tungsten-halogen white-light lamp. {\bf b}, Dielectric permittivity measured
at photon energy $h\nu = 0.6$ eV upon cooling (green points) and warming
(blue points). The hysteresis curve represents switching between the antiferromagnetic
insulating G-AFM-I and paramagnetic metallic PM-M phases with positive and
negative permittivity values, respectively. Both phases coexist as stripes
inside the hysteresis loop. {\bf c}, Snapshots of
stripe formation from the videos in the Ancillary files recorded
during slow cooling ($\lesssim 3 $ K/min, left panel) and fast cooling ($\simeq
5.2$ K/min, right panel). Bright regions correspond to the G-AFM-I phase
and dark regions to the PM-M phase. During fast cooling, additional metastable
stripes are detected, as evident from regions of stronger dark contrast,
highlighted by the red box. {\bf d}, Illustration of the reversible photoinduced
switching between G-AFM-I and PM-M phases. The numbers and corresponding
frames represent different moments along the timeline of the recorded video
in the Ancillary files, and at different locations on the sample
(frames 1 to 3 located at the blue box and frames 4 to 7 at the green box
in {\bf a}). In frame 1, captured under background light with intensity of
$\sim 0.2$ mW/cm$^2$, the coexisting phases are stabilized by temperature
at the red dot in {\bf b}. Following the increase of excitation light irradiance
reaching the sample surface to above the threshold of $\sim 1$ mW/cm$^2$,
the PM-M phase rapidly expands over the full macroscopic extent of the sample
(frames 2 to 4) and the entire crystal enters the supercooled metallic phase
indicated by the end of the red arrow in {\bf b}. When the white light intensity
is reduced to the background intensity before time 5, the insulating G-AFM-I
phase spontaneously reappears at the opposite end of the sample and then
the original phase configuration of the sample is restored (frames 7 and
1).}
\label{fig:light}
\end{figure*} \clearpage

To clearly visualize the photoinduced switching behavior, we stabilize the
stripe pattern present in frame 1 of Fig. 1d at $\sim$ 54 K while cooling
under low background illumination, which corresponds to the red point in
Fig. 1b. When white light illumination is increased above a threshold intensity
of only $\sim 1$ mW/cm$^2$, the existing dark PM-M stripes rapidly expand
to cover the entire sample with a characteristic stripe domain wall velocity
of $0.3 $ mm/s. The resulting thermally inaccessible metallic state persists
as long as the light illumination remains above the threshold
irradiance. When the light is reduced to the background level, the recovery
of the exact initial stripe domain morphology occurs over ten seconds and
begins with spontaneous nucleation  of the G-AFM-I domain at the opposite
end of the sample, which exhibits re-entrant behavior. To effectively capture
a complete view of the photoinduced switching process we provide real-time
recorded videos in the Ancillary files.

The stripe domain structure and its light-induced development signify the
strong effect of elastic strain on the phase transition. The transition to
the insulating phase is accompanied by disproportionate changes in the crystal
lattice parameters $\Delta b/b = 1.23 \%$ and $\Delta c/c = -0.97 \%$, with
a minute change in $a$, $\Delta a/a = 0.05 \%$ \cite{Krautloher2018}. The
concomitant spontaneous elastic strain is adapted by the formation of interfacial
domain walls along the $a$ axis, extending obliquely deep into the bulk of
the sample \cite{McLeod2021}. When the illumination intensity exceeds the
threshold, we find an avalanche-like enlargement of the metal domains with
the successive conversion of the insulating domains due to essentially coherent
motion of the domain walls throughout the entire crystal volume. This is
in contrast to the manganites, where the PIPT remains  confined near the
illuminated region \cite{Fiebig1998,Zhang2016,McLeod2020}.

In the following, we develop an assessment of the intrinsic instabilities
in the electronic, magnetic, and structural properties of \CRTO \ to place
the light switching behavior into full context. We discuss these instabilities
and their manifestations in the low energy electrodynamics, including the
phonon and interband transition spectra of \CRTO. We argue that the optically
driven cooperative phenomena result from collective redistribution of electrons
within the $\rm Ru$ $4d$ orbital manifolds, which governs the structural
distortions of $\rm RuO_6$ octahedra and spin structures of $\rm Ru$ moments.

\section*{Lattice distortions across the insulator-to-metal transition}

\begin{figure*}[t]
\includegraphics[width=16cm]{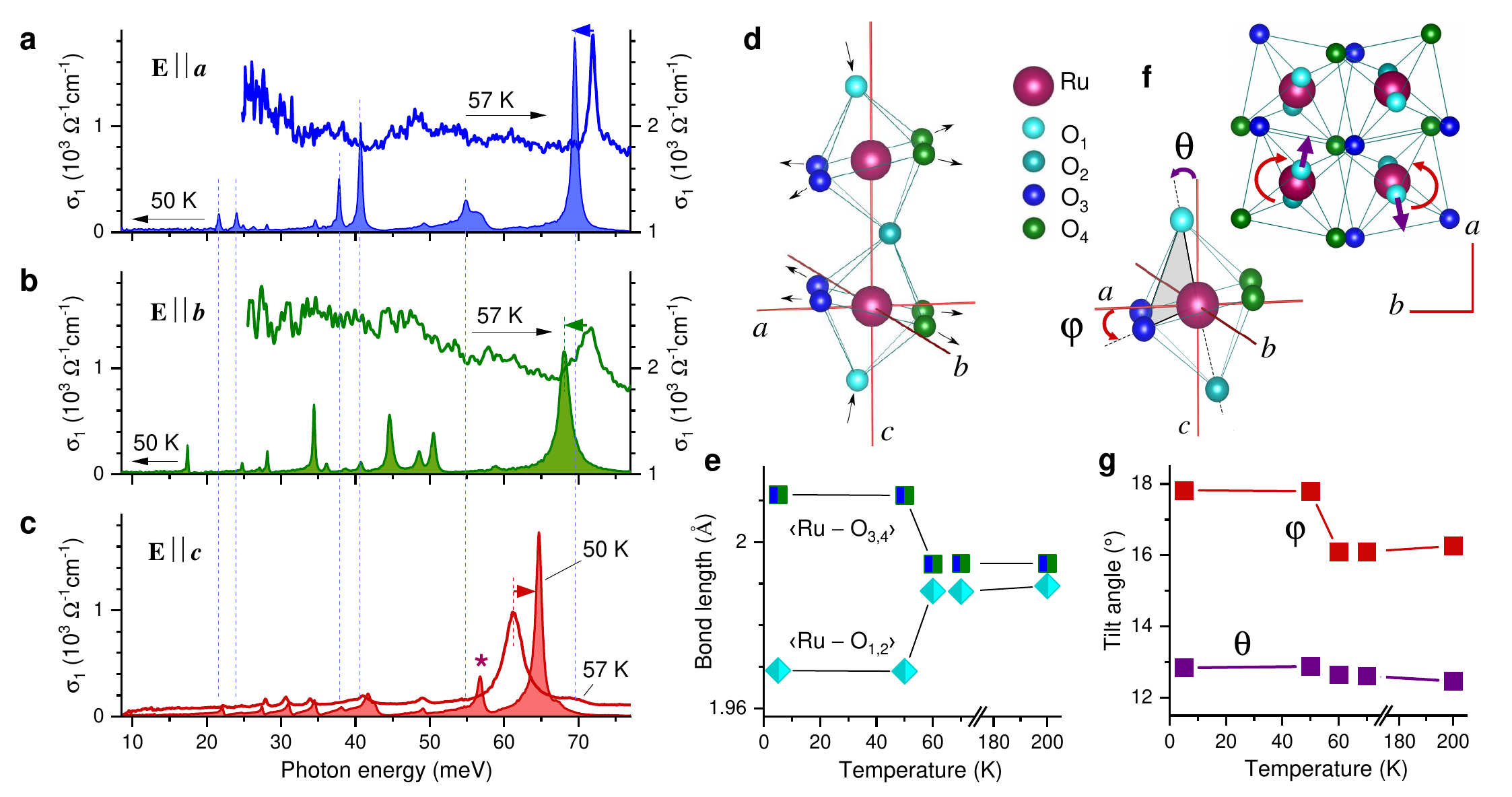}
\caption{\scriptsize{\bf Phonon spectra evidence for crystal instability.\
}\scriptsize
{\bf a-c}, Changes in the phonon spectra along the $a$, $b$, and $c$ crystallographic
axes above and below the IMT ($T_{IMT}$ = 55 K). {\bf d-g}, An illustration
of the octahedral distortions across the isostructural transition. The distortions
include changes in the average equatorial $\langle $Ru--O$_{3,4}$$\rangle
$ and apical $\langle $Ru--O$_{1,2}$$\rangle$  distances ({\bf d, e}) along
with rotations of the RuO$_6$ octahedra expressed in azimuthal $\varphi$
and polar $\theta$ angles for the octahedron diagonal connecting the apical
O$_1$ and O$_2$ oxygens ({\bf f, g}). $\varphi$ and $\theta$ correspond to
the $X^+_2$ octahedral rotation and $X^-_3$ octahedral tilt, respectively.
The electronic background present above the transition disappears below $T_{IMT}$,
exposing a number of infrared active phonons. The vertical dotted lines in
{\bf a-c} serve as guides to mark the eigenfrequencies of the most pronounced
$a$-axis phonon modes. Arrows indicate the softening of the in-plane stretching
modes ({\bf  a, b}) and the hardening of the out-of-plane vibrational mode
({\bf c}) consistent with the compression of the RuO$_6$ octahedra along
the $c$-axis, as illustrated in panels {\bf d} and {\bf e}. The rise of the
mode marked by the asterisk in panel {\bf c} is possibly related to the concomitant
rotation and tilting of the RuO$_6$ octahedra, as illustrated in panels {\bf
f} and {\bf g}.}
\label{fig:phonons}
\end{figure*}

The concomitant changes of the electronic transport and crystal structure properties across $T_{IMT}$ are captured by the evolution of the infrared phonon spectra, as illustrated in Fig. 2. While the electronic background in the metallic state obscures most in-plane phonons, this background is rapidly suppressed below $T_{IMT}$, revealing the presence of a number of infrared active modes in the insulating phase. From factor group analysis, there are 19 $A_1$, 19 $B_1$, and 17 $B_2$ zone-center infrared active phonon modes that should be observable in the $a$-, $b$-, and $c$-axis spectra, respectively, within the space group $Bb2_1m$ \cite{Smith2019}. We have identified most of these phonon modes and list their frequencies, linewidths, and oscillator strengths in Table I of Appendix F. The most intense and highest frequency mode, corresponding to Ru-O bond stretching, exhibits a shift in frequency in accordance with the change in the bond distances (Fig.~2d,e) that occurs as a result of the $c$-axis RuO$_6$ octahedral compression. In the transition, the $c$-axis compression and octahedral distortions are much stronger than in the parent \CROb \ compound. 

As in pristine bilayer \CROb, the metallic state above $T_{IMT}$ is 
quasi-two-di\-men\-sional with strong uniaxial anisotropy in the electronic transport \cite{Yoshida2004,Lin2005}. The low background in the $c$-axis IR optical conductivity allows well-defined phonons to be retained across the transition. In addition to the marked shift of phonon frequencies due to changes in the Ru-O bond distances, changes in octahedral orientation (Fig.~2f,g) also lead to specific variations in the phonon spectrum. The increase of these orthorhombic distortions gives rise to phonon features that are not active in the aristotype tetragonal $I4/mmm$ structure. In particular, some of the phonon eigenvectors overlap with a single symmetry-adapted mode of the irreducible representation of $I4/mmm$, which represents either the $X^+_2$ rotation of the RuO$_6$ octahedra around the $c$-axis or the $X^-_3$ diagonal tilting mode \cite{Smith2019}. This behavior is demonstrated by the rise of the phonon  mode marked by the asterisk at 57~meV in Fig.~2c. The observed high sensitivity of the mode intensity is expected for excitations that transform primarily as the $X^+_2$ and $X^-_3$ irreducible representations, which drive the transition to the polar $Bb2_1m$ phase \cite{Benedek2011}.

\section*{Mott gap excitations and competing orders}

The extent of the opening of the optical gap and associated transfer of spectral weight upon cooling is shown by the broadband (far-IR to UV) optical conductivity $\sigma(\omega)$ and dielectric permittivity $\varepsilon(\omega)=\varepsilon_1(\omega)+i\varepsilon_2(\omega)=1+4\pi
i\sigma(\omega)/\omega$ presented in Fig.~3. In this spectral range the response along the $a$ and $b$ axes does not exhibit appreciable in-plane anisotropy. Above $T_{IMT}$ the PM-M state shows a clear in-plane free-charge-carrier response with negative $\varepsilon_1(\omega)$ below 1 eV, consistent with the corresponding permittivity value in Fig.~1b.
The metallic response gives an effective carrier density per Ru atom of $n_{eff}^{D} = 2m/\pi e^2N_{Ru}\times\omega^{2}_p/8 \approx 0.4 \ e^{-}/\mathrm{Ru}$, where $m$ is the free electron mass, $\omega_p\approx 2.8$ eV is the plasma frequency, and $N_{Ru}=1.38\times 10^{22}$ cm$^{-3}$. Below $T_{IMT}$ a clean optical gap of $E_g^{dir} = 0.45$ eV opens with a narrow Urbach tail and the in-gap spectral weight, $SW(\Omega)=\int_0^{\Omega}\sigma_1(\omega)d\omega$, shifts to energies as high as 5-6 eV. This in-gap spectral weight, $n_{eff} = 2m/\pi e^2N_{Ru}\times SW(E_g^{dir})\approx 0.2 \ e^{-}/\mathrm{Ru}$, accounts for as much as half of $n_{eff}^{D}$ (see Fig.~3d). The out-of-plane response in Fig.~3e,f, on the other hand, does not display free-charge-carrier behavior, but rather a broad peak centered near 1.2 eV gives rise to the low background in $\sigma_1(\omega)$ at phonon frequencies above $T_{IMT}$ in Fig.~2c, suggestive of incoherent interlayer hopping in quasi-two-dimensional \CRTO. Upon cooling through $T_{IMT}$, $\sigma_1^c$ undergoes significant changes in a similar way as $\sigma_1^{a}$, including the opening of the gap and spectral weight shift to higher energies across the entire spectral range.

\begin{figure*}[t]
\includegraphics[width=16cm]{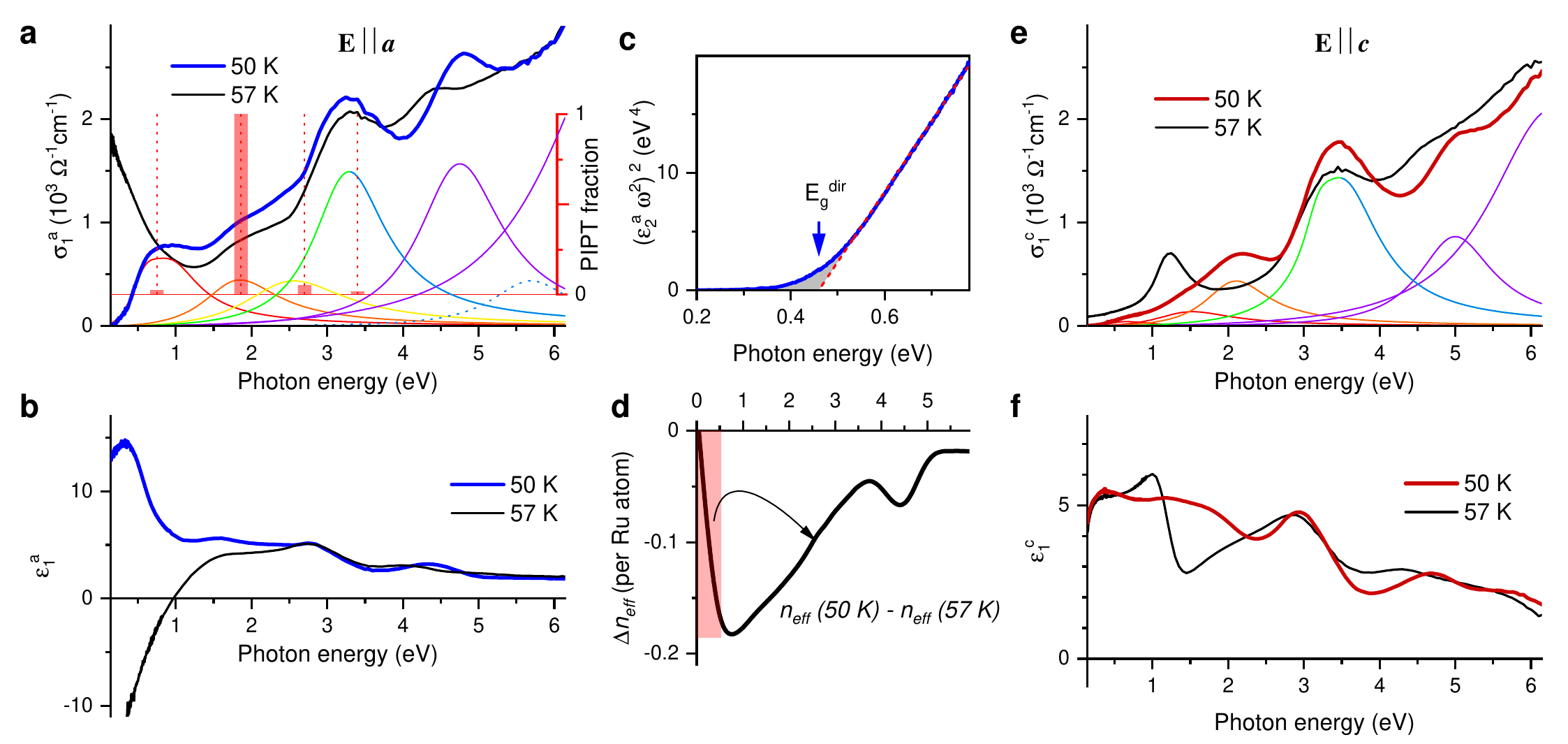}
\caption{\scriptsize{\bf Optical gap and spectral weight transfer across the insulator-to-metal
transition.\ }\scriptsize
{\bf a-b}, Real part of the in-plane optical conductivity ({\bf
a}) and dielectric permittivity  ({\bf b}) measured above (thick black) and
below (thick blue) $T_{IMT}$. {\bf c}, Tauc plot evaluated using the $a$-axis
dielectric function. The linear fit (red dotted line) gives the direct band
gap $E_g^{dir}\approx 0.45$ eV. The small sub-gap absorption (gray shaded
area) corresponds to minor contribution from Urbach tail states. {\bf d},
Redistribution of spectral weight upon cooling obtained from integrating
the difference of the spectra in {\bf a} in terms of the effective number
of electrons per Ru atom, $n_{eff}$. The in-gap spectral weight (red shaded
area) transfers over large energy scales $\sim$ 5 eV above the gap, pointing
to strong electronic correlation effects. {\bf e, f}, Real part of the optical
conductivity and dielectric permittivity measured along the $c$-axis above
(thick black) and below (thick red) $T_{IMT}$. The thin colored lines in
{\bf a} and {\bf e} correspond to separate interband transitions determined
by dispersion analysis of the low temperature spectra. The red bars in {\bf a} indicate the volume fraction of the PIPT under constant low irradiance of 20 $\mu$W/cm$^2$ of monochromatic light at the selected photon energies.}
\label{fig:SW}
\end{figure*}

We note that even though 1\% Ti substitution critically changes the properties of the ground state of the system, this has almost no effect on the optical conductivity spectra of the metallic phase. The electronic structure of pristine \CROb, including its manifestations in optical properties, can be captured well by density functional theory (DFT) band structure calculations \cite{Markovic2020,Singh2006} (for optical conductivity calculations see Appendix Fig.~7). The severe changes in the spectra across the Mott metal-insulator transition can also be addressed in a straightforward way by taking into account the on-site Coulomb repulsion $U$ within the Ru $d$ shell, which competes with the kinetic energy on the order of the Ru $t_{2g}$ bandwidth $W$, and results in Hubbard-like band splitting. In order to explain the observed anomalies and the anisotropy of the optical response, spectra obtained from relativistic DFT$+U$ calculations for the experimental crystal structure, assuming G-AFM order, are compared with the measured optical spectra along the $a$- and $c$-axes below $T_{IMT}$ (see Fig.~4a and 4b, respectively). In contrast to DFT results, the DFT$+U$
solution is insulating, as evidenced by the partial densities of states (PDOS) in Fig.~4c. A Mott gap separates empty Ru $d_{xz\downarrow,yz\downarrow}$
from occupied $d_{xy\downarrow}$ states. This orbital ordering within Ru
$t_{2g\downarrow}$ states is stabilized due to the compression of RuO octahedra
below $T_{IMT}$. With the exchange parameter $J_H=0.8$ eV, $U = 2.8$ eV is selected by matching the calculated absorption peaks to the experimental spectra, which are decomposed into individual bands in Fig.~3a,e by a simultaneous fit of a sum of Lorentzians to $\sigma_1(\omega)$ and $\varepsilon_1(\omega)$ (see Table II in Appendix F). The direct band gap is found to be $\sim 0.6$ eV, fairly consistent with the experimental value. The experimental $E_g^{dir} = 0.45$ eV can be reproduced by decreasing $U$ to 2.5 eV, which still adequately describes the optical transitions. 

To identify the optical transition responsible for the PIPT, we illuminate
the sample with monochromatic light at selected photon energies near the
peak positions in Fig. 3a. Only the optical band peaked at 1.85 eV exhibits
resonance behavior, where the PIPT is triggered under constant irradiation
with average laser intensity as low as 20 $\mu$W/cm$^2$ following the same
dynamics shown in Fig. 1d and the Ancillary videos. What is the specific nature of
this resonance?
The origin of the absorption bands in Fig.~3a,e is elucidated
by comparison with theoretical spectra decomposed into additive contributions
calculated as transitions between non-overlapping ranges of initial and final
bands (see Fig.~4a,b), which are assigned through the analysis of the PDOS in Fig.~4c. Two absorption bands, lying at 0.8 eV and 2.5 eV (red and yellow peaks), are assigned to be due to weakly allowed Ru $4d$ intersite transitions from the occupied $d_{xy\downarrow}$ to the unoccupied $d_{xz\downarrow,yz\downarrow}$ upper Hubbard band and the $e_{g}$ orbitals, respectively. These transitions
are critically sensitive not only to $U$ but also to light polarization and Ru magnetic order within a bilayer, and become almost completely suppressed in the $c$-axis spectra. On the other hand, the band lying at 1.85 eV (orange peak), consisting of transitions from majority $t_{2g\uparrow}$ to minority $d_{xz\downarrow,yz\downarrow}$, remains largely unchanged between the $a$- and $c$-axis spectra. The best overall agreement with the experimental spectra is obtained for G-AFM order, where neighboring Ru ions are aligned antiferromagnetically within the bilayer. Figure 4d shows the calculated spectra based on the assumption of an alternative AFM-$b$ magnetic ordering, where Ru ions are aligned ferromagnetically within the bilayer, for comparison. The main absorption peak at  $\sim 1.85$ eV in Fig.~4a,b is completely suppressed in Fig.~4d as $d_\uparrow \longrightarrow d_\downarrow$ transitions between majority and minority spin-polarized Ru
$4d \ t_{2g}$ bands become forbidden, restricted by the Pauli principle.
We assign the PIPT resonance band at 1.85 eV to intersite transitions between
neighboring sites $i$ and $j$ of the form $t^{4}_{2g}(i)$ $t^{4}_{2g}(j)
\longrightarrow t^{3}_{2g}(i)$ $t^{5}_{2g}(j)$. In the final low-spin $S=1/2$
local excited state, the electron is transferred to the unoccupied $d_{xz}$
or $d_{yz}$ orbital on the neighboring Ru site with anti-aligned magnetic
moment. 

\begin{figure*}[t]
\includegraphics[width=16cm]{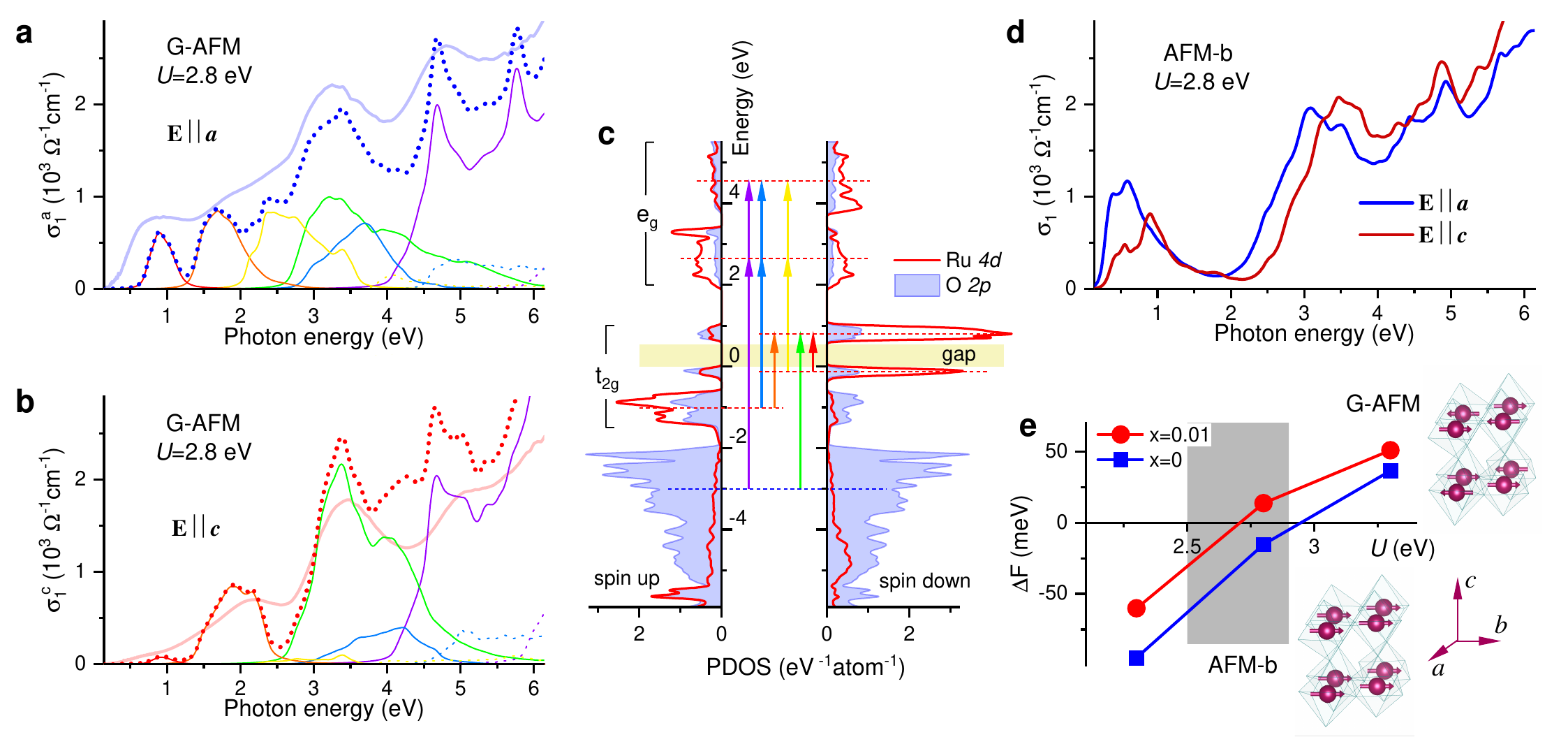}
\caption{\scriptsize{\bf Spin-controlled Mott-Hubbard bands and electronic phase instability.\ }\scriptsize
{\bf a-b}, Real part of the in-plane ({\bf a}) and out-of-plane ({\bf b})
optical conductivity calculated by DFT$+U$ assuming G-AFM magnetic order
(thick dotted lines) with a breakdown into separate orbital contributions
(thin colored lines). The low-temperature orthorhombic crystal structure
of \CRTO \ is used.  The on-site Coulomb repulsion $U=2.8$ eV is chosen to
match the positions of the peaks in the corresponding experimental spectra
(semitransparent thick lines and {\bf Fig. 3 a, e}). {\bf c}, Partial densities
of the majority spin (spin up, left panel) and minority spin (spin down,
right panel) Ru $4d$ (red lines) and oxygen $2p$ (blue shaded areas) states.
Colors of the vertical arrows denote the orbital character of the separate
optical bands in {\bf  a} and {\bf b} and in {\bf Fig. 3 a, e}. {\bf d},
Uniaxial anisotropy of the optical conductivity calculated by DFT$+U$ assuming
AFM-b magnetic order. The crystal structure and the on-site Coulomb repulsion
$U$ are the same as in {\bf a} and {\bf b}. The ferromagnetic alignment of
the Ru moments within the bilayers leads to the suppression of transitions
between majority and minority spin-polarized Ru $4d \ t_{2g}$ bands (orange
peaks in {\bf a} and {\bf b} and arrow in {\bf c}) with a shift of the associated
spectral weight to the lower energy minority band transitions below 1 eV.
{\bf e}, Energy gain between the G-AFM and AFM-$b$ states per formula unit
as a function of $U$ for two experimental crystal structures of \CRTOx, $x=0$
(blue) and $x=0.01$ (red). The range of values of $U$ consistent with {\bf
a} and {\bf b} (gray area) corresponds to the degeneracy of the magnetically
ordered G-AFM and AFM-$b$ states illustrated by the images of Ru moments
(red)surrounded by oxygen octahedra.}
\label{fig:DFTU}
\end{figure*}

Our first-principles DFT$+U$ calculations based on experimental structural parameters strongly describe the light-polarization-dependent and spin-controlled low-energy electrodynamics of \CRTO \ by incorporating the value of $U$ of 2.5 to 2.8 eV. Moreover, the calculations also reveal a critical electronic phase instability with respect to the magnetic order and crystal structure distortions. We calculated the energy difference between G-AFM and AFM-$b$ magnetic orders as a function of $U$ for the crystal structures of pristine \CROb\ and \CRTO\ (see Fig. 4e). For small $U$ values, a relatively large gain in kinetic energy favors the AFM-$b$ ground state with FM order within bilayers, while stronger electronic correlations stabilize the G-AFM phase. Critically, in the range of values of $U$ defined above (gray shaded area) these two states are nearly degenerate. By comparing these results (red circles) to the same calculations for the crystal structure parameters of the parent compound (blue squares), we find that the structural distortions addressed in Fig.~2d-g also play in favor of the insulating ground state with AFM-ordered nearest-neighbor Ru moments. The relative stability of the G-AFM ground state is achieved by tuning the system through a Mott transition by only 1\% replacement of Ru by Ti, which reduces the effective Ru electronic bandwidth $W$ while retaining proximity to the collective instabilities with a high degree of susceptibility to external stimuli. A metastable AFM-$a$ or $b$ phase may manifest itself within the thermal hysteresis due to transient trapping upon fast cooling (see Fig.~1c). More significantly, the considered intrinsic
instabilities in the electronic, magnetic, and structural properties of \CRTO
\ give rise to phase switching triggered by the $d_{\uparrow} \rightarrow
d_{\downarrow}$ transitions.

\section*{Discussion}

The peculiar character of the PIPT excitation band peaked at 1.85 eV differs
from all other absorption bands in that it involves the specific concomitant
change of both the spin state of neighboring Ru atoms ($S=1\rightarrow S=1/2$)
and the local orbital polarization ($t_{2g} \longrightarrow d_{xz,yz}$),
intertwining the charge, spin, and orbital degrees of freedom. The upper
bound on the critical density of local excited states is determined by the
threshold photon flux of $7\times 10^{13}$ s$^{-1}$cm$^{-2}$. Photoinduced
expansion of the metallic phase across the full extent of the sample volume
requires an irradiation time of $10-30$ s, giving an estimation of the total
photon fluence needed to switch the entire sample that is consistent with
the planar (not bulk) density of Ru atoms, $N^{(ab)}_{Ru}=6.6\times 10^{14}$
cm$^{-2}$. This low fluence suggests that bulk switching corresponds to at
least $10^5-10^6$ Ru states changed per quantum of light absorbed. Such avalanche
behavior points to the cooperative interaction between Ru sites photoexcited
locally at the interface between the two phases, which mediates the pump-induced
motion of the interface and macroscopic expansion of the metallic phase domains
\cite{Koshihara2022}.

The observed dynamics of photoinduced expansion of one phase with respect
to another resembles that recently addressed by coupled first- and second-order
time-dependent Ginzburg-Landau parameters \cite{Sun2020}. Applied to the photoinduced IMT in manganites \cite{McLeod2020}, a strain-coupled Ginzburg-Landau theory has been considered on the basis of coupled order parameters $Q$ and $M$, denoting the amplitude of a dominating Jahn-Teller $\rm MnO_6$ octahedral distortion and the  ferromagnetic moment, respectively. A similar approach can also underlie the description of the photoinduced phase transition in \CRTO, which is caused by the interplay between the structural $\rm RuO_6$ distortions and competing AFM and FM spin structures of intra-bilayer $\rm Ru$ moments. The non-centrosymmetric polar $Bb2_1m$ structure dictates some specific features of the structural and magnetic order parameters. First, examining the effect of structural distortions on the phase behavior of isostructural improper ferroelectric insulators $\rm Ca_2Ti_3O_7$ and $\rm Ca_2Mn_3O_7$ reveals that the structural order parameter $Q$ in this class of bilayered perovskites is related to the distortion amplitudes for the individual rotation ($Q_{X_2^+}$) and tilt ($Q_{X_3^-}$) modes that primarily drive the transition to the $Bb2_1m$ from the aristotype $I4/mmm$ phase \cite{Benedek2011,Senn2015}. The hybrid order parameter $Q_{X_{23}} = Q_{X_2^+}Q_{X_3^-}$ defines minima in the double-well potential of the total energy landscape around the reference $I4/mmm$ structure. The same octahedral distortions that produce the polar phase also couple to the magnetic ordering. Second, in contrast to manganites, the itinerant FM state in \CRTO \ cannot be described by a single magnetic order parameter $M$ because two distinct sublattices of ferromagnetically ordered $\rm Ru$ bilayers, $\mathbf{M}_{\rm I}$ and $\mathbf{M}_{\rm II}$, are antiferromagnetically coupled and modulated along the $c$-axis. The phenomena of metamagnetic texture in pristine \CROb \ has been described by the Ginzburg-Landau theory for the specific coupling between the two order parameters $\mathbf{l}=1/2(\mathbf{M}_{\rm I}-\mathbf{M}_{\rm II})$ and $\mathbf{f}=1/2(\mathbf{M}_{\rm I}+\mathbf{M}_{\rm II})$, corresponding to the antiferromagnetic staggered magnetization and ferromagnetic spin polarization, respectively \cite{Sokolov2019}. We argue that the above specific features are the key ingredients building the resulting flattened energy landscape in \CRTO, whose exceptionally small perturbations by light cause switching between G-AFM-I and PM-M phases within the thermal hysteresis loop.

Another distinct difference from the case of manganites is the coherent insulator-metal domain wall propagation throughout the entire crystal volume under the action of light. The minimum required dilute substitution of Ru in \CRTO \ needed to establish the Mott insulating ground state leaves the crystal quality almost intact. As a result, there is no marked pinning of the domain walls, and the hysteresis loop in Fig. 1b remains essentially symmetrical (see e.g. Ref.\cite{Fan2011} for comparison). Increasing the substitution level leads to increased disorder with stronger pinning effects and nanoscale fragmentation of the stripe configuration, and the regular Ginzburg-Landau phenomenological approach becomes no longer sufficient to describe the pump-induced phase interface motion \cite{Sun2020}. 

More critically, increasing the substitution of $\rm Ru$ ions further reduces the effective $\rm Ru$ electronic bandwidth $W$ so that the kinetic energy no longer competes with the on-site $U$. The critical temperature $T_{IMT}$ (which is 55 K for $x=0.01$ and falls between the two critical temperatures of the pristine \CROb, $T_N$ = 56 K and $T_s$ = 48 K \cite{Krautloher2018,Bao2008}, see Appendix Fig.5c) becomes larger for higher $x$, reaching $\approx $ 95 K at $x =0.1$ \cite{McLeod2021}, signifying the stabilization of the Mott insulating state. The collective instabilities in the electronic, magnetic, and structural properties of these compounds and their manifestations in the charge dynamics as discussed above thereby become insensitive to manipulation by light.

\section*{Summary and outlook}

We have reported here a unique Mott insulator state in \CRTO, achieved through
delicate control of the one-electron bandwidth $W$ by dilute isovalent substitution
of Ru. By combining comprehensive optical measurements (terahertz to UV)
with spin-polarized DFT and DFT$+U$ calculations we parameterize this state
and assign Hubbard bands associated with the observed optical transitions.
The corresponding IMT exhibits exceptional sensitivity to external stimuli
such that local low-fluence resonant photoexcitation of Ru $t_{2g}$ $d_\uparrow
\longrightarrow d_\downarrow$ transitions triggers avalanche-like switching
of the entire macroscopic sample to the metallic phase. Moreover, dilute
substitution maintains the lattice structure and keeps the crystal quality
intact, which is distinct for bandwidth- and filling-controlled Mott IMTs
in correlated oxides. From a fundamental perspective, the elimination of
pinning effects enables coherent photoinduced motion of the insulator-metal
interface and makes \CRTO \ an ideal model system for building and testing
a theory of Mott transition dynamics in the presence of cooperative electron-electron
and electron-lattice interactions. At the same time, from a technology perspective,
the intact crystal quality and low-fluence light sensitivity pave the way
to possible new designs for nanodevices that achieve quantum-level photosensitivity.

\begin{acknowledgments}
%\begin{acknowledgements}

This project was supported by the European Research
Council under Advanced Grant No. 669550 (Com4Com). We gratefully acknowledge
P.~Radhakrishnan, L.~Wang, P.~Puphal for XRD, F.~Predel for SEM-EDX, S.~Hammoud
for ICP-AES, and R.~K.~Kremer for specific heat measurements. High-resolution
neutron diffraction experiments were performed at the SPODI instrument operated
by FRM II at the Heinz Maier-Leibnitz Zentrum (MLZ), Garching, Germany.
%\end{acknowledgements}
\end{acknowledgments}

\subsection*{APPENDIX A: Sample preparation and characterization.}

High-quality single crystals of \CRTO \ were grown using an optical floating-zone technique. Energy dispersive X-ray (EDX) analysis and inductively coupled plasma atomic emission spectroscopy (ICP-AES) verified the uniform sample stoichiometry, and \CRTO \ crystal structure parameters were derived as a function of temperature based on high-resolution neutron diffraction data. The derived lattice parameters in space group $Bb2_1m$ change across the transition at $T_{IMT}=55$ K from  $a = 5.3685$ \AA, $b=5.5979$ \AA, and $c=19.3478$ \AA \ at 50 K to $a = 5.3659$ \AA, $b=5.5295$ \AA, and $c=19.5359$ \AA \ at 60 K. Neutron diffraction experiments confirmed a pure G-type AFM phase below $T_{IMT}$. Details of crystal growth and characterization, including crystal and magnetic structure determination, are available in Ref. \cite{Krautloher2018}. With special care, only single-domain crystals with characteristic dimensions of $2\times2\times0.2$ mm$^3$ were selected based on initial examination via reflectivity contrast by polarized-light optical microscopy. Measurements of the magnetization on every selected sample were performed using a vibrating sample magnetometer (see Appendix Fig.~5). The single crystals were oriented using backscattering Laue X-ray diffraction (XRD) and supplementary high-resolution XRD measurements with a four-circle setup. The XRD results confirm the monodomain nature of our selected samples, which is further evident in the measured phonon spectra in Fig. 2 a-c. Samples were cleaved prior to optical measurements. Intrinsic properties could only be obtained for cleaved surfaces, as mechanical treatment such as polishing introduced uncontrolled artifacts due to possible surface damage and strain. Polishing also led to a dramatically broadened transition and shifted $T_{IMT}$, reflecting the sensitivity of electronic phase instabilities discussed in the main text.

\subsection*{APPENDIX B: Spectroscopic ellipsometry.}

We used broadband spectroscopic ellipsometry to measure the complex dielectric
function, $\varepsilon (\omega) = \varepsilon_1 (\omega)+i\varepsilon_2 (\omega)=1+4\pi
i[\sigma_1 (\omega)+i\sigma_2 (\omega)]/\omega$, over a range of photon energies
extending from the far infrared ($\hbar\omega = 0.01$ eV) into the ultraviolet
($\hbar \omega = 6.5$ eV). The $a$- ($b$- or $c$-) axis component
of the dielectric tensor $\varepsilon_a$ ($\varepsilon_b$ or $\varepsilon_c$)
corresponds to the measured pseudodielectric function $\varepsilon_{a,b,c} \approx \tilde \varepsilon_{a,b,c}$ at angle of incidence ranging from 70$^{\circ}$ to 80$^{\circ}$ for sample orientations with the $a$ ($b$ or $c$) axis in the plane of incidence. The pseudodielectric function  $\tilde \varepsilon$ is derived by a direct inversion of the ellipsometric parameters $\Psi$ and $\Delta$ assuming bulk isotropic behavior of the sample surface. For details of data acquisition and analysis in the case of optically anisotropic crystals, see Refs.\cite{Larkin2017,Larkin2018}.
In the frequency range 7.5 meV to 1 eV we used home-built ellipsometers in combination with Bruker IFS 66v/S and Vertex 80v Fourier transform infrared spectrometers. The measurements in the far infrared (7.5 to 88 meV) utilized synchrotron edge radiation of the 2.5 GeV electron storage ring at the IR1 beamline of the Karlsruhe Research Accelerator (KARA) at the Karlsruhe Institute of Technology, Germany. The measurements in the range 0.6 eV to 6.5 eV were performed with a Woollam variable angle ellipsometer of rotating-analyzer type.

\subsection*{APPENDIX C: First-principles calculations.}
The relativistic band structure calculations were performed using the linear muffin-tin orbital (LMTO) method as implemented in PY LMTO computer code \cite{Antonov2004}. PBESol exchange-correlation potential was used \cite{Perdew2008}. The Coulomb interaction of Ru $4d$ electrons in the presence of spin-orbit coupling (SOC) was taken into account using the rotationally invariant DFT+$U$ method \cite{Yaresko2003}. The interband contribution to the imaginary part of the dielectric tensor was calculated using the dipole approximation to the matrix elements of the momentum operator \cite{Antonov2004}. For the calculations, we used the experimental structural data according to Ref.\cite{Krautloher2018}.

\subsection*{APPENDIX D: Visualization and control of the photoinduced phase transition.}
A reflected-light microscopy setup equipped with a He flow optical cryostat
and either a tungsten-halogen white light lamp or monochromatic laser sources
of selected photon energies were used to excite and record the PIPT in \CRTO.
Representative videos and screenshots  are available. The bulk and resonant character of the phase switching under low fluence irradiation compared with specific heat data (Appendix Fig.~6) allows us to exclude any light-induced heating effects.

\subsection*{APPENDIX E: Supplementary figures.}

\begin{figure*}[h]
\includegraphics[width=16cm]{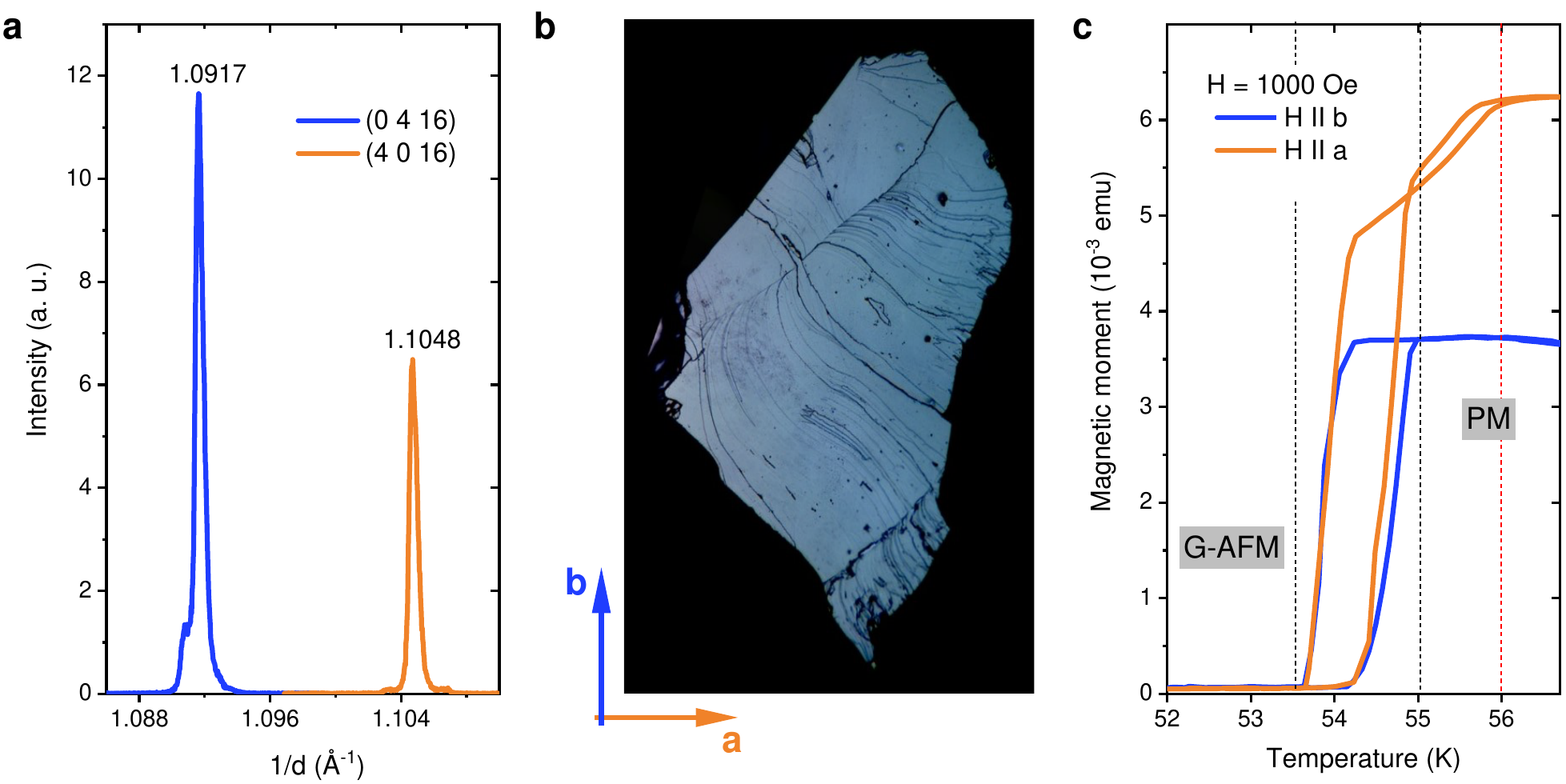}
\caption{\scriptsize{\bf Twin-free \CRTO.\ }\scriptsize
{\bf a}, X-ray diffraction pattern of \CRTO \ near (0 4 16) and (4 0 16) peaks. {\bf b}, Polarizing microscope image under crossed polarizers in reflectance mode. {\bf c}, Temperature dependence of the magnetic moment measured in the external field of $H = 1000$ Oe, with $H\ \|\ a$ (orange) and $H\ \|\ b $ (blue). These results confirm the monodomain nature of the representative sample. The vertical black dotted lines reperesent the width of the hysteresis loop in Fig. 1b. The red dotted line denotes the AFM- a magnetic ordering temperature $T_N = 56$ K in pristine \CRO.
}
\label{fig:XRD}
\end{figure*}
\clearpage

\begin{figure*}[t]
\includegraphics[width=13cm]{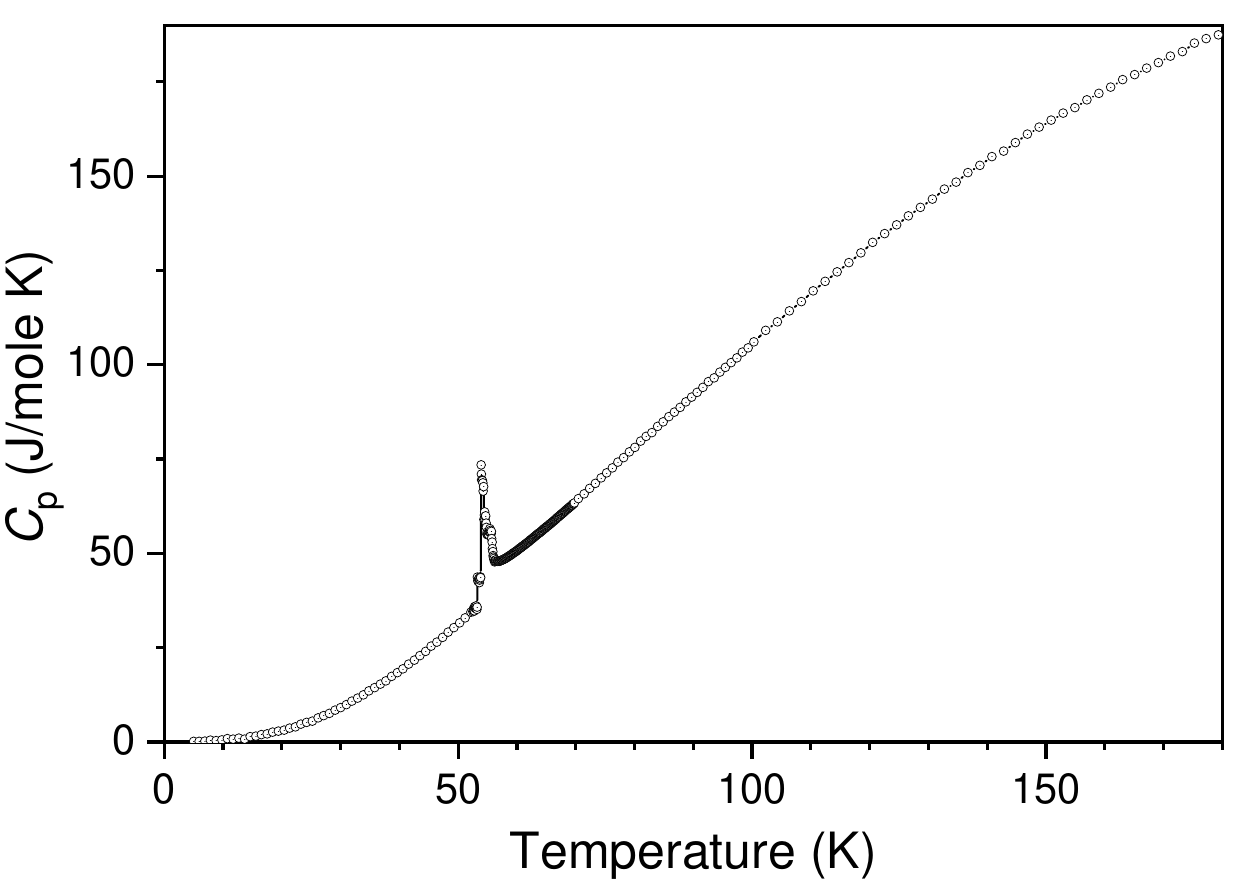}
\caption{\scriptsize{\bf Temperature dependence of the specific heat.
}\scriptsize
Specific heat was measured in the temperature range between 2 and 180 K with a Physical Properties Measurement System (Quantum Design) using the thermal relaxation technique. A low ramp rate of 5 mK/min was used in the vicinity of the IMT. The PIPT is triggered under constant irradiation
with average 1.86 eV laser intensity as low as 20 $\mu$W/cm$^2$. The total time necessary to fully switch the sample is on the order 10-30 s, which corresponds to an upper bound of $S\cdot $500 $\mu$J on the amount of energy transferred to the sample surface $S$. In the absence of thermal exchange and accounting for the Cp peak maximum of less than 100 J/mol\ K, this upper bound corresponds to a change in sample temperature of $\Delta T < 10^{-3}$ K. The cryostat cold finger coupling ensures that $\Delta T$  remains more than an order of magnitude smaller than this value.  This estimate, along with the bulk and resonant character of the PIPT, allows us to exclude any light-induced heating effects.
}
\label{fig:Cp}
\end{figure*}

\begin{figure*}[t]
\includegraphics[width=16cm]{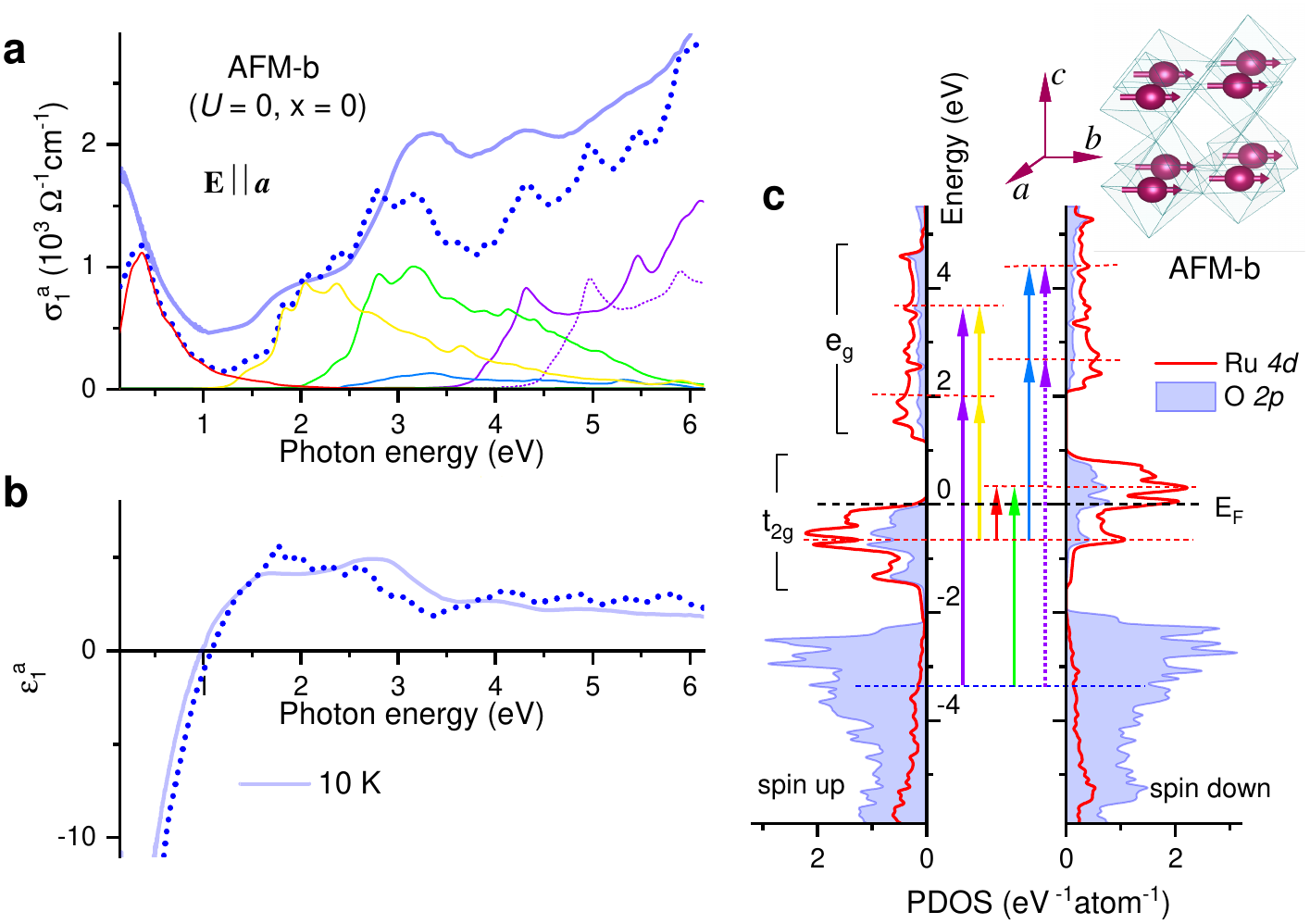}
\caption{\scriptsize{\bf DFT assignment of interband transitions in \CRO.\ }\scriptsize
{\bf a-b}, Real part of the $a$-axis optical conductivity ({\bf a}) and dielectric permittivity ({\bf b}) calculated by DFT assuming AFM-$b$ magnetic order (thick dotted lines) with a breakdown into separate orbital contributions. The low-temperature orthorhombic crystal structure of pristine \CRO \ is used. The corresponding experimental spectra of \CRO \ are shown (semi-transparent thick blue lines). {\bf c}, Partial densities of the majority spin (spin up, left panel) and minority spin (spin down, right panel) Ru $4d$ (red lines) and oxygen $2p$ (blue shaded areas) states. Colors of the vertical arrows denote the orbital character of the separate optical bands in {\bf a} and {\bf b}.
}
\label{fig:DFT}
\end{figure*}

\begin{figure*}[h]
\includegraphics[width=15cm]{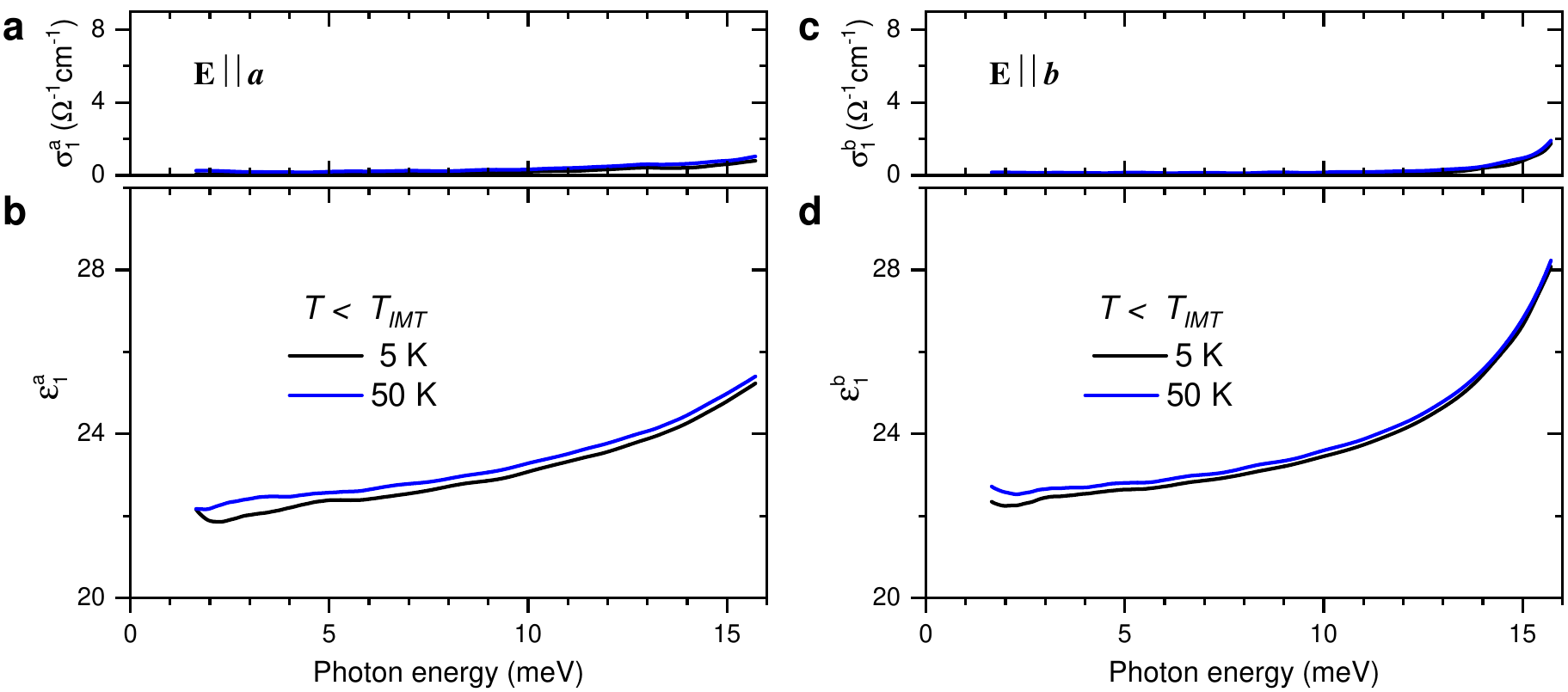}
\caption{\scriptsize{\bf THz dielectric response.\ }\scriptsize{\bf a-b}, Terahertz spectra of \CRTO \ measured by time-domain THz transmission spectroscopy at temperatures below the IMT. The real part of the optical conductivity ({\bf a}) and dielectric permittivity ({\bf b}) as measured along the $a$-axis. {\bf c-d}, the same as measured along the $b$-axis. The spectra confirm the absence of impurity related in-gap absorption.
}
\label{fig:THZ}
\end{figure*}\clearpage

\subsection*{APPENDIX F: Supplementary tables.}
\begin{table}[!h]
\centering
\begin{scriptsize}
\caption{\scriptsize {\bf Phonon mode parameters.\ }
The observed phonon modes in \CRTO for the $a$-, $b$-, and $c$-axis polarizations. A minimal set of $N$ Lorentzian oscillators was used to simultaneously fit the far-infrared complex dielectric function. The resulting fitting parameters are the resonance frequency $\omega_0$, oscillator strength $\Delta \epsilon$, and width $\gamma$. }
\begin{tabular}{ cccc|ccc} 
 \hline
 \hline
  & & a-axis (50K) & & & b-axis (50K)\\
N &$\omega_{0}(meV)$&$\Delta \epsilon$ & $\gamma (meV)$ &$\omega_{0}(meV)$&$\Delta \epsilon$ & $\gamma (meV)$\\
 \hline
1 & 21.52 & 0.51 & 0.10 & 17.48 & 0.74 & 0.04\\ 
2 & 23.96 & 0.72 & 0.25 & 20.76 & 0.09 &  0.36\\ 
3 & 26.40 & 0.17 & 0.40 & 24.79 & 0.15 & 0.12\\ 
4 & 28.08 & 0.08 & 0.10 & 27.18 & 0.10 & 0.15\\ 
5 & 34.65 & 0.20 & 0.25 & 28.12 & 0.38 & 0.19\\ 
6 & 37.79 & 0.90 & 0.33 & 34.43 & 1.18 & 0.25\\ 
7 & 40.78 & 2.00 & 0.40 & 36.07 & 0.22 & 0.39\\ 
8 & 49.51 & 0.26 & 1.55 & 38.69 & 0.17 & 0.70\\ 
9 & 54.97 & 0.76 & 1.30 & 40.70 & 0.36 & 0.77\\ 
10 & 56.53 & 0.47 & 1.58 & 44.66 & 1.26 & 0.63\\ 
11 & 61.61 & 0.07 & 1.71 & 48.48 & 0.48 & 0.77\\ 
12 &   &   &   & 50.49 & 0.73 & 0.65\\
13 &   &   & & 58.92 & 0.11 & 0.97\\ 
 &   &   &  &   &   & \\
Ru-O& stretching& mode& 50K (57K)& &\\ 

14 & 69.60 & 1.88 & 0.67 & 68.11 & 2.87 & 1.62  \\ 
 & (72.55) & (2.17) & (1.90) & (70.86) & (4.78) & (3.27) \\ 
 &   &   &  &   &   &  \\
  \hline
 & & c-axis (50K) & & & c-axis (57K)\\
 N &$\omega_{0}(meV)$&$\Delta \epsilon$ &$ \gamma (meV)$ &$\omega_{0}(meV)$&$\Delta \epsilon $& $\gamma (meV)$\\
  \hline
1 & 12.92 & 0.22 & 0.40 & - & - & -\\
2 & 22.07 & 0.39 & 0.28 & 22.09 & 0.12 & 0.55\\
3 & 24.72 & 0.05 & 0.40 & 24.76 & 0.04 & 0.25\\ 
4 & 27.41 & 0.36 & 0.39 & 27.87 & 0.27 & 0.45\\ 
5 & 30.96 & 0.44 & 0.45 & 30.64 & 0.38 & 0.73\\
6 & 34.50 & 0.47 & 0.47 & 33.93 & 0.36 & 1.17\\
7 & 38.24 & 0.06 & 0.45 & 39.17 & 0.22 & 1.80\\
8 & 41.82 & 0.80 & 0.95 & 41.22 & 0.78 & 1.98\\
9 & 42.64 & 0.16 & 0.56 & - & - & -\\
10 & 49.14 & 0.11 & 0.63 & 48.96 & 0.48 & 2.38\\
11 & 56.83 & 0.66 & 0.73 & - & - & -\\
 &   &   &  &   &   &  \\
Ru-O &stretching&mode& & &\\
12 & 64.68 & 2.85 & 0.92 & 61.20 & 5.10 & 3.00 \\
 &   &   &  &   &   &  \\
13 & 67.21 & 0.21 & 1.66 & 68.81 & 0.24 & 3.51\\
 \hline
  \hline
\end{tabular}
\end{scriptsize}
\end{table}

\begin{table}
\centering
\begin{scriptsize}
\caption{\footnotesize {\bf Optical band parameters.\ } 
The fitting parameters (resonance frequency $\omega_0$, oscillator strength $\Delta \epsilon$, and width $\gamma$) of separate Lorentzian bands determined by simultaneous fit to the optical conductivity of and dielectric permittivity 50 K spectra of \CRTO in Fig. 3a-b and e-f.
}
\begin{tabular}{ cccc } 
 \hline
 &  &  &  \\
  &$\omega_{0}(eV)$&$ \Delta \epsilon$ & $\gamma (eV)$\\
  &  &  &  \\
 \hline
  &  &  &  \\
$a$-axis &  &  & \\ 
   & 0.623 & 4.66 & 0.553\\ 
  & 1.000 & 2.56 & 0.811\\
  & 1.850 & 1.18 & 1.212\\
 & 2.563 & 0.81 & 1.633\\
  & 3.290 & 1.20 & 1.171\\ 
  & 4.752 & 0.70 & 1.347\\ 
  & 5.690 & 0.13 & 1.290\\
  &  &  &\\
 $b$-axis  &  &  & \\ 
  & 0.623 & 6.22 & 0.553\\ 
 & 1.000 & 2.74 & 0.878\\
  & 1.850 & 1.14 & 1.083\\
 & 2.563 & 0.81 & 1.568\\
  & 3.338 & 1.26 & 1.241\\ 
  & 4.704 & 0.77 & 1.429\\ 
 & 5.798 & 0.08 & 1.123\\
   &  &  &\\
 $c$-axis &  &  & \\ 
   & 0.665 & 0.49 & 0.669\\ 
 & 1.530 & 0.54 & 1.217\\
  & 2.108 & 0.76 & 1.037\\
  & 3.190 & 0.12 & 0.449\\ 
 & 3.539 & 0.87 & 1.132\\
  & 4.999 & 0.31 & 1.199\\ 
 \hline
\end{tabular}
\end{scriptsize}
\end{table}

\subsection*{APPENDIX G: Ancillary Videos description.}

{\bf Video1Aslow.mov}: 

The nucleation process of mixed-phase states within the hysteresis loop across the first order metal-to-insulator phase transition in \CRTO, upon cooling at a slow rate of $\lesssim 3$ K/min. Dark regions correspond to the paramagnetic metallic (PM-M) phase, and bright stripes to the insulating G-AFM phase.
\\

{\bf Video1Bfast.mov}: 

The same nucleation process as depicted in Video1Aslow, but recorded at a fast cooling rate of $\simeq 5.2$ K/min. Additional stripes of stronger dark contrast are detected (see, for example, the lower portion of the sample at timestamp 00:05), which represent possible transient trapping of a metastable ferromagnetic phase (AFM-$a$ or -$b$).

{\bf Video2PIPT.mov}: 

This video illustrates the reversible switching between insulating and paramagnetic metallic phases by light, recorded at $T = 54$ K within the hysteresis loop. The sample is initially in the mixed state under the background light irradiance of $\sim 0.5$ mW/cm$^2$, with the region marked by the blue square in Fig.1a displaying clear insulating/metallic stripes (clock starts, 00:00). When the white light illumination is increased above the threshold intensity of $\sim 2$ mW/cm$^2$ (00:04), the stripe domain walls move and the metallic phase rapidly expands over the full extent of the sample. When the light is once again reduced to the background intensity (00:23), the insulating G-AFM-I phase spontaneously reappears at the opposite end of the sample and then the original phase configuration of the sample is restored. 

%\bibliography{RefCa3Ru2O7PIPT2022Apr}
%\end{document}

%

\end{document}